\documentclass{aa}
\usepackage{graphicx}
\usepackage{txfonts}
\newcommand{\kms}{\mbox{km~s$^{-1}$~}}
\begin{document}
\title{The peculiar galaxy IC~1182: An ongoing merger?\thanks {Based on data obtained with the Nordic Optical telescope (La Palma, Spain) with ALFOSC. Also based on observations obtained with FORS1 at the ESO Very Large Telescope (Paranal, Chile), and on observations collected with DFOSC at the D1.54m telescope at the European Southern Observatory (La Silla, Chile)} 
}
              \author{
                 M. Moles\inst{1}
            \and D. Bettoni\inst{2}
            \and G. Fasano\inst{2}
            \and P. Kj\ae rgaard\inst{3}
            \and J. Varela\inst{1}
                                  }

\offprints{M. Moles}

                            \institute  {
                  Instituto de Astrof\'{\i}sica de Andaluc\'{\i}a,
                  Consejo Superior de Investigaciones Cient\'{\i}ficas\\
                  Apdo. 3004, E-18080 Granada \\
                  \email:moles@iaa.es, jesusv@imaff.cfmac.csic.es
        \and Osservatorio Astronomico di Padova,
              Vicolo Osservatorio 5, I-35122 Padova \\
              \email: bettoni@pd.astro.it, fasano@pd.astro.it
        \and  Copenhagen University Observatory. The Niels Bohr Institute for
                  Astronomy Physics and \\ Geophysics, Juliane Maries Vej 30, DK-2100 Copenhagen \\
              \email: per@astro.ku.dk
                         }

\date{Received 9 September 2003 / Accepted  19 January 2004}
\authorrunning{Moles et al.}
\titlerunning{The Peculiar Galaxy IC1182}

\abstract {High resolution broad and narrow band images and long slit spectroscopy of the peculiar galaxy IC~1182 are  presented. The analysis of the broad band images  reveals a distorted morphology with a large, heavily obscured disk-like structure and several knots in the central region. Galactic material, some of it in the form of two slender tails, is detected well beyond the main body of the galaxy. The second, fainter tail and several knots are reported here for the first time. The galaxy has color indices of an early type object except U$-$B, which is significantly bluer than what is typical for this kind of galaxy. 

The narrow band images centered on different emission lines show that the galaxy is a very powerful emitter. Most of  the knots detected in the central region and in the prominent tail emerging eastward from the galaxy are very luminous in H$\alpha$, and have typical sizes about 1 kpc (FWHM). The emission in the main lines extends all over the galaxy, with plumes and arc-like structures seen in H$\alpha$ at large distances from the center. The observed, uncorrected H$\alpha$ flux corresponds to a total luminosity of 3.51$\times$10$^{41}$ erg s$^{-1}$, about 3 times that of the starburst galaxy Arp~220. We have found that the internal extinction deduced from the observed Balmer decrement is high all along the slit, with E$_{B-V}\approx$ 1, so the corrected SFR could amount to 90~M$_{\odot}$ per year. On this basis IC~1182 is found to be a very powerful starburst galaxy. Surprisingly, the source is not in the IRAS Point Source Catalogue.

The emission knots detected in the central region of the galaxy have line ratios that place them close to the border of the region occupied by active nuclei in the diagnostic diagrams. Using the best determined diagnostic ratio, [OIII]/H$\beta$ {\sl vs}  [NII]/H$\alpha$, they can still be classified as extreme HII-like regions. We notice that the same kind of line ratios are also measured at different places in the galaxy, adding to the idea that the nuclear line ratios can be explained in terms of stellar photoionization. The metallicity we have measured for the ionized gas in the two brightest central knots is low, 0.1~Z$_{\odot}$ and 0.06~Z$_{\odot}$ respectively, and their measured helium abundance is also lower than solar.

In the main body of the galaxy, besides the reported knots, the distribution of the ionized gas resembles that of an inclined disk about 12 kpc in size. The spectroscopic data show however a complex rotation pattern. We interpret them  as corresponding to two identifiable disk galaxies with observed rotation amplitudes of 200 \kms and 100 \kms respectively. The stellar absorption lines detected in the bigger system do not show any clear rotation pattern.

The data presented here indicate that IC~1182 is a high luminosity starburst system. Its global properties and peculiarities can be understood as corresponding to two systems that can still be recognized, in the process of merging, with two tidal tails emerging from the central region of the galaxy. In the main tail there are several candidates forming tidal dwarf galaxies. The measured low metallicity of the ionized gas, together with the low amplitude of one of the systems, suggests that the process involves a late-type, gas-rich spiral galaxy that is supplying most of the gas to the system.


         }

\maketitle

\section{Introduction}

IC~1182 (Mkr~298) is a bright, early-type (S0p) galaxy belonging to the Hercules cluster. It is placed at the end of the stream of galaxies running from the center of the cluster (defined by its brightest galaxies, NGC~6041A, B) toward the East. Its peculiar morphology is apparent on DSS images, where a long, linear, knotty structure extending 88\arcsec~ from the main body of the galaxy to the E is clearly visible. In the picture by Arp (1972) diffuse material associated with the galaxy is also visible toward the NW. These are clear signs of interaction. The impression that the galaxy is involved in some interaction or merging process is reinforced by the discovery reported by Rafanelli et al. (1999) of  two bright knots separated by $\approx$~3\farcs3, at PA $\approx$ 130 in the central region of the galaxy.

IC~1182 has some other peculiarities. Bothun et al. (1981) found that it has total colors that correspond to an early type object, except U$-$B, that is too blue for an E or S0 galaxy. They also reported (see also  Salpeter  \& Dickey 1985) that IC~1182 has an unusually large HI mass fraction for an early type galaxy. Their data reveal the contribution to the observed HI profile of a close companion, most likely interacting with IC~1182, but even after taking into account that contamination, IC~1182 is still very HI-rich, with log(M$_{HI}$/L$_B)\approx-$0.5. On the other hand, IC~1182 is not associated with a powerful radio-continuum source. Dickey and Salpeter (1984) reported the  detection at 1415~MHz of a rather weak source, probably extended, at the position of IC~1182.

Another sign of peculiarity of IC~1182 is the sizeable polarization it shows (p = 1.09\%, at $\theta$= 104.5\degr, Martin et al. 1983). It is also an  X-ray source. Huang and  Sarazin (1996) reported detection of an individual source coincident with IC~1182 when making high resolution ROSAT observations of the Hercules cluster. Rafanelli et al. (1999) pointed out that the soft X-ray luminosity of the nucleus of IC~1182 is close to the mean value for Seyfert 2 galaxies. They also argued that the existing data on the X-ray variability of the source would imply a source size too small to contain the supernovae necessary to produce the observed luminosity. This would be strong evidence for the presence of an active nucleus at the center of IC~1182.

The nature of the nucleus of IC~1182 has been the subject of some discussion. It was first classified as Seyfert 2 by Khachikian and Weedman (1974), but the line ratios reported by Koski (1978) are closer to those of a LINER (see Viegas-Aldrovandi  \&  Gruenwald 1990). Heckman et al. (1981) reported that the observed line profiles are due to the superposition of several components that are spatially and kinematically distinct, and V\'eron et al. (1997) classify it as a composite spectrum object.

The reported properties and the high internal extinction and high H$\alpha$ luminosity we find in the present work (see below) should correspond to a high IR luminosity galaxy. Surprisingly, the only IR source detected by IRAS at the position of IC~1182 is one close to the bright knot at the end of the long tail emerging toward the East of
the main body of the galaxy, with  f$_{60}$ = 240~mJy in the IRAS PSC (Beichman et al. 1986), corresponding to L$_{FIR}\sim$ 10$^7$ L$_{\odot}$ at the distance of the galaxy.

We present here new photometric and spectroscopic data for IC~1182 gathered with the NOT (La Palma), the  1.54m  Danish telescope (La Silla), and the VLT Antu (U1, Paranal). These include broad and narrow band images under good to excellent seeing conditions, long slit spectra with three different slit orientations, and MOS observations of the detected knots. The new data reveal the presence of several knots in a highly obscured central region, a second, faint tail almost perpendicular to the main one, a complex rotation pattern of the ionized gas, and evidence for a generalized star formation process all over the galaxy. These peculiarities lead us to classify IC~1182 as a starburst galaxy, resulting from an ongoing merging process involving two spiral galaxies, one of them gas rich and of rather late type.

\section{Observations and data reduction}

The log of the photometric observations is given in Table~\ref{Log}. We made three short exposures of IC~1182 with the direct CCD camera attached to the NOT in May~95, two through the Gunn~r filter and the third in Bessel~B. The spatial scale is 0\farcs188/pixel. The seeing conditions were exceptional, 0\farcs40 and 0\farcs53, respectively. Inspection of the co-added image of the two red exposures led us to notice a linear feature that, starting in the inner region of the galaxy, runs northward, almost perpendicular to the previously known eastward optical tail. That feature could not be appreciated in the noisier B image. This prompted new observations that were carried out in 1998 and 1999.
%
\begin{table}
      \caption[]{Log of  the Photometric  Observations} 
\label{Log} 
$$
      \begin{array}{lcrrcc} 
\hline \noalign{\smallskip} 
{\rm Filter} & {\rm WL} & {\rm FWHM} & {\rm ExpT} & {\rm Date} & {\rm Seeing} \\ & nm & nm~~~~ & s~~~~ & & \\
     \noalign{\smallskip}   \hline \noalign{\smallskip} 

{\rm Gunn~r} & 680.0  & 102.0 & 2\times120  &{\rm 05/95} & 0\farcs40 \\
{\rm Bes~ B}  & 440.0 & 100.0 &  120 &{\rm 05/95} & 0\farcs53 \\ 
{\rm Bes~ B} & 440.0 & 100.0  & 2\times1000 &{\rm 08/98} & 1\farcs05 \\  
{\rm Bes~ V} &  530.0 & 80.0 & 2\times700 &{\rm 08/98} &  1\farcs15  \\ 
{\rm  Gunn~r} &  680.0 & 102.0   & 2\times500 &{\rm 08/98} & 0\farcs80 \\ 
{\rm Bes~ U} & 362.0 & 60.0 & 600 &{\rm 08/99} & 0\farcs89 \\ 
{\rm  Bes~ U} & 362.0 & 60.0 & 1800 &{\rm 08/99} & 1\farcs10 \\ 
{\rm Bes~ B} & 440.0 & 100.0 & 300 &{\rm 08/99} & 0\farcs91 \\ 
{\rm Bes~ V} & 530.0 & 80.0 & 300 &{\rm 08/99} & 0\farcs76 \\  
{\rm Gunn~ r} & 680.0 & 102.0 & 600 &{\rm 08/99} & 0\farcs80 \\ 
{\rm Bes~ I} & 797.0 & 157.0 & 300 &{\rm 08/99} & 0\farcs61 \\ 
{\rm Bes~ I} & 797.0 & 157.0 & 600 &{\rm 08/99} & 0\farcs80 \\  
$[OIII]$ & 521.4 & 10.5 & 1800 &{\rm 08/99} & 0\farcs76 \\ 
{\rm Strg~ y} & 547.0 & 22.0 & 1500 &{\rm 08/99} & 0\farcs73 \\  
{\rm H\alpha~ N} & 678.8 & 4.5 & 1200 &{\rm  08/99} & 0\farcs66 \\ 
{\rm H\alpha~ W} & 657.7 & 18.0 & 1200 &{\rm  08/99} & 0\farcs76 \\ 
$[SII]$ & 696.5 & 5.0  & 300 &{\rm 08/99}  &  0\farcs65 \\ 
$[SII]$  & 696.5 & 5.0 & 2400   &{\rm 08/99} &   1\farcs31 \\ 
$[SII]$  & 696.5   &    5.0  & 1800    &{\rm   08/99}  &   0\farcs92  \\
         \noalign{\smallskip} \hline

        \end{array} $$

             \end{table}

All subsequent images were taken with ALFOSC. The standard Bessel U, B, V, I, and  Gunn r filters were used   for the broad band images. We also obtained narrow band filter images, centered on the (redshifted) [OIII]$\lambda$5007, H$\alpha$, and [SII]$\lambda\lambda$6717,6731~lines, and continuum images trough filters centered in nearby spectral regions (Str\"omgen y for the [OIII] line, and H$\alpha$ wide for the H$\alpha$ and [SII] lines). In all cases standard stars for extinction and calibration purposes were observed.

The reduction process of the broad band images was done using the standard IRAF packages\footnote{IRAF is the Image Analysis and Reduction Facility made available to the astronomical community by the National Optical Astronomy Observatories, which  are operated by the Association of Universities for Research in Astronomy (AURA), Inc., under contract with the U.S. National Science Foundation.}. The galaxy IC~1182 is included in our long term program of photometry of galaxies in nearby clusters and the photometric calibration of the broad band images was done following the steps described  in Fasano et al. (2002). The internal calibration errors are smaller than 0.03 magnitudes in all the bands. Once calibrated, the package AIAP (Fasano  1990) was used to produce accurate models of the galaxy. Its capability to create any kind of interactive, {\sl  ad~hoc} masking of the images makes it particularly well suited to analyze the     existence of sub-structures. The program is used to generate faithful, smooth versions of the overall light distribution  of the galaxy avoiding the influence of the possible embedded features. The model galaxy is then subtracted from the  observed image, leaving the  structural features, if any, very distinctly shown.

The narrow band, line and continuum images were also reduced and calibrated using the IRAF packages. Before subtracting the continuum, the line and continuum images were background subtracted, shifted and rotated until perfect matching between both images, and smoothed to the same effective seeing. This was performed using the IRAF task HJCOMP (Hansen and J{\o}rgensen 1998), that results in an accuracy better than 0.1 pixel, or 0\farcs02 in our case. The continuum level was fixed in such a way that normal E galaxies in the field (D77 in our case) disappeared when  subtracted. This could be   done at the 5\% level. The calibration was established using spectrophotometric standard stars observed through the same filters. 

Unavoidably, some of the narrow filters include nearby emission lines that can influence the final result. For example, the [NII]$\lambda\lambda$6548,6584 lines are within the wavelength range covered by the H$\alpha$ line filter. The H$\alpha$ wide filter is also contaminated, this  time by the [OI] lines. Their importance can be judged from the line ratios derived from the spectra. We find that it amounts to 19\% of the signal in the line filter, and to 4\% in the
continuum filter, with small variations from point to point. The [OIII]$\lambda$5007 line filter includes the other nebular line, [OIII]$\lambda$4959. This is however not very important since this line enters only the wing of the filter.  On the other hand, the Str\"omgren y filter we used to get the continuum produces ghosts and the background is not completely smooth. In fact, this is the limiting factor in detecting faint [OIII] features. However, whereas that contamination affects the flux level and the detection of very faint features in the [OIII] line, it does not change the conclusions about the general distribution and ionization structure of the emitting gas. Taking all these effects into account, we have estimated that the final accuracy in the total flux we reached is better than 5\% in H$\alpha$, and  better than 10\% in [OIII] and [SII], as judged from the results for the standard stars.

The spectroscopic data were collected with three different instruments: ALFOSC, attached to the NOT (La Palma), DFOSC attached to the 1.54m Danish telescope at La Silla, and FORS1 attached to VLT Antu (UT1) at Paranal (see     Table~\ref{LogS}). Data reduction and calibration were done using the IRAF packages. Intermediate resolution spectra were taken with grisms \#7 (covering from 3700\AA~ to 6800\AA) and \#8 (covering from 4700\AA~ to 7200\AA), using two similar
spectrographs, DFOSC and ALFOSC. We found that both sets of data are very similar in all respects. Given that the ALFOSC data have better sampling and spatial resolution, and a higher S/N, we will refer only to them in the following. The spatial sampling was 0\farcs376/pixel. We also compared the two  spectra taken with grisms \#7 and \#8 in the common region. The matching is better than 10\% in the continuum flux level, and much better in the line ratios, so the
data were merged in one set covering from [OII]$\lambda$3727 to the [SII]$\lambda\lambda$6717,6731 lines. Comparison  of the two spectra is indicative of the flux errors in our data. 

We have also compared the H$\alpha$ flux measured in our spectra with the flux measured in the narrow band image for the same area. We find a satisfactory agreement, within 30\%, between the two measurements. 

The zero point of  the wavelength calibration was controlled with the sky lines, producing a nominal accuracy of 0.2\AA.  The MOS exposure was intended to get information on the redshift and excitation ratios along the main tail. The data were reduced and calibrated in a similar way.

A higher resolution spectrum was taken with ALFOSC using grism \#13. The slit at PA = 113 was placed through the center of the galaxy and one of the bright knots, denoted CKN2 (see Figure~\ref{centralknots}). The stellar velocity dispersion values were obtained with the Fourier Quotient technique. We also obtained a higher resolution spectrum with the VLT ANTU unit, using FORS1 and grism ~600R. The slit was placed through the knots CKN2 and CKN4 (see Figure~\ref{centralknots}). The flux calibration was based on observations of the spectrophotometric standard star LTT~7987. The emission lines were measured by fitting Gaussian profiles. 
%
\begin{table}
      \caption[]{Log of the Spectroscopic Observations} 
\label{LogS}
      $$   
\begin{array}{lrrrrrr}   
\hline  \noalign{\smallskip}  
{\rm Telescope} & {\rm Date~~~} & {\rm  Grism~~~} & {\rm Slit} & {\rm PA}  & {\rm \AA/px} & {\rm ExpT(s)} \\  \noalign{\smallskip} \hline \noalign{\smallskip}

{\rm ALFOSC} & {\rm 08/99} &  7 &  1\farcs2 & 128  & 1.48 & 1800  \\ 
             & {\rm 08/99} &  8 &  1\farcs2 & 128 & 1.24 & 1800   \\
             & {\rm 03/00} & 13 &  1\farcs2 & 113 & 0.52 &  2400  \\ 
{\rm ALFOSC/MOS} & {\rm 03/00} & 7 & 1\farcs2 &   -- & 1.48  &  1800 \\  
{\rm DFOSC}  &  {\rm 08/99} &  7 & 1\farcs2 & 128 & 1.48 & 1800 \\ 
             & {\rm  08/99} & 8  & 1\farcs2 & 128 & 1.24 & 1800  \\ 
{\rm VLT/FORS1}  & {\rm 08/99}  & 600R& 0\farcs8& 62 & 1.05 & 300 \\ 

\noalign{\smallskip} \hline
\end{array}
 $$ \end{table}
%
%

\section{Overall morphology of IC~1182 and embedded structures}

\begin{figure*}
\centering
\includegraphics[angle=-90, width=\textwidth]{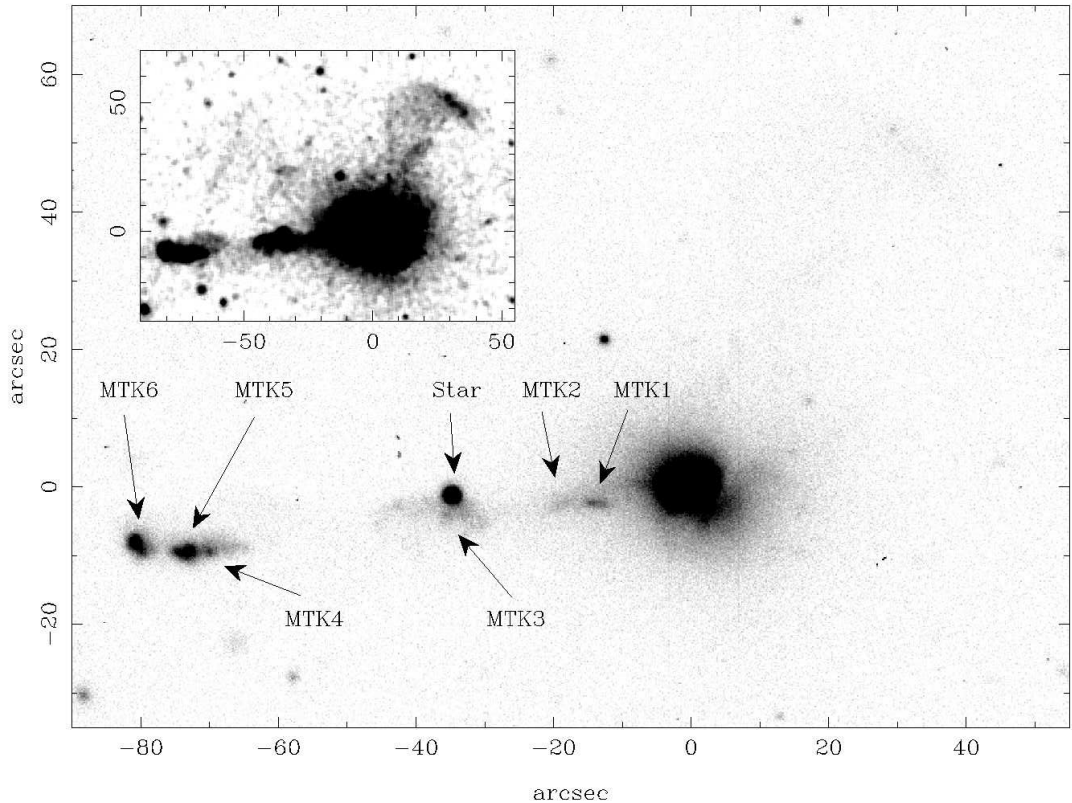}
\caption{B-band image of IC~1182. North up, East left. The plot is centered in CK. Notice the extension of the material around the galaxy toward the NW, and the absorption feature in the central body. In the upper left insert the same  image, median filtered, is presented to enhance the fainter features}
\label{Bfilter}
\end{figure*}

The well-exposed broad band images of IC~1182 have been used to measure the global properties of the galaxy (see Figure~\ref{Bfilter}). The main photometric parameters of IC~1182 that we have obtained are given in Table~\ref{magIC1182}. They are corrected for galactic extinction, that amounts to A$_B$ = 0.218 mag (Schlegel et al. 1998). The B magnitude and the B$-$V and V$-$R color indices we find are in good agreement with the cataloged values, whereas our U$-$V is about 0.3 mag bluer. The absolute B magnitude given in the Table (we have used H$_0$ = 70 \kms Mpc$^{-1}$) is also corrected for internal extinction, that we have estimated from the Balmer decrement measured from the long slit spectra. As we will see below, a representative value of the color excess for the gas in the main body of the galaxy is E$_{B-V}$ = 1. Using the recipe given by Calzetti (1997) for star forming galaxies, this corresponds to 0.44 for the starlight, which we have applied to IC~1182. The total, corrected luminosity we find amounts to M$_B = -$22.67. This makes IC~1182 a very luminous galaxy. With the new, corrected, M$_B$ value, the ratio M(HI)/L(B) is 15 times lower than that reported by Bothun et al. (1981).
%

In the upper left insert in Figure 1, we have plotted the median filtered image of IC~1182 to show all the extensions, including the structure toward the NW. A faint, knotty, linear feature emerging from the galaxy main body toward the North is also visible. The most salient aspect revealed by the broad band images in the main body of the galaxy is a large scale absorption feature in the SW side (see Figure~\ref{Bfilter}). Otherwise, the overall aspect of the galaxy is close to an spheroidal system, not too dissimilar to an early type  object. Indeed, parameters such as the ellipticity, $\varepsilon$, or the position angle, PA, are not well defined, in the sense that they change with the radius, in particular not too far from the center. But when the analysis is restricted to the outer regions, from r $\approx$ 12\farcs2 outward, those parameters are better defined and, if not exactly constant, they change in a narrow range, with $\varepsilon\approx$~0.25, and PA~$\approx$~80. These values are similar in all bands.
%
\begin{table}
     \caption[]{Main  properties of IC 1182$^a$} 
\label{magIC1182} 
$$
   \begin{array}{lr} 
\hline \noalign{\smallskip} 
{\rm B } & 15.18 \\
  {\rm U-B} & -0.16 \\ 
{\rm B-V} & 0.98 \\ 
{\rm V-Gr}& 0.31 \\ 
{\rm  V-I} & 1.16 \\ 
{\rm logf_{H\alpha}~(erg/cm^2/s)} & -12.87 \\ 
{\rm
     \varepsilon^b} & 0.25 \\ {\rm PA^b} & 80\degr \\
\hline
{\rm   cz^c  (\kms)} & 10300   \\  {\rm  M_B}  &   -22.67 \\ 
{\rm  log L_{H\alpha} (erg s^{-1})} & 42.71 \\ 
{\rm log (L_B/M_{HI})} & 0.67 \\
\noalign{\smallskip}
         \hline \end{array} 
$$
\begin{list}{}{}
\item[a] 
Apparent magnitudes and the H$\alpha$ flux are corrected for external extinction only, with A$_B$ = 0.218. M$_B$ and the H$\alpha$ luminosity are also corrected for internal extinction, as explained in the text. A value of H$_0$ = 70 km/s/Mpc and a redshift of 10300 \kms have been used to estimate the distance 
\item[b] Values for the external isophotes of the galaxy
\item[c] The redshift corresponds to CK
\end{list}
\end{table}
\begin{figure*}
\centering
\includegraphics[angle=-90,width=\textwidth]{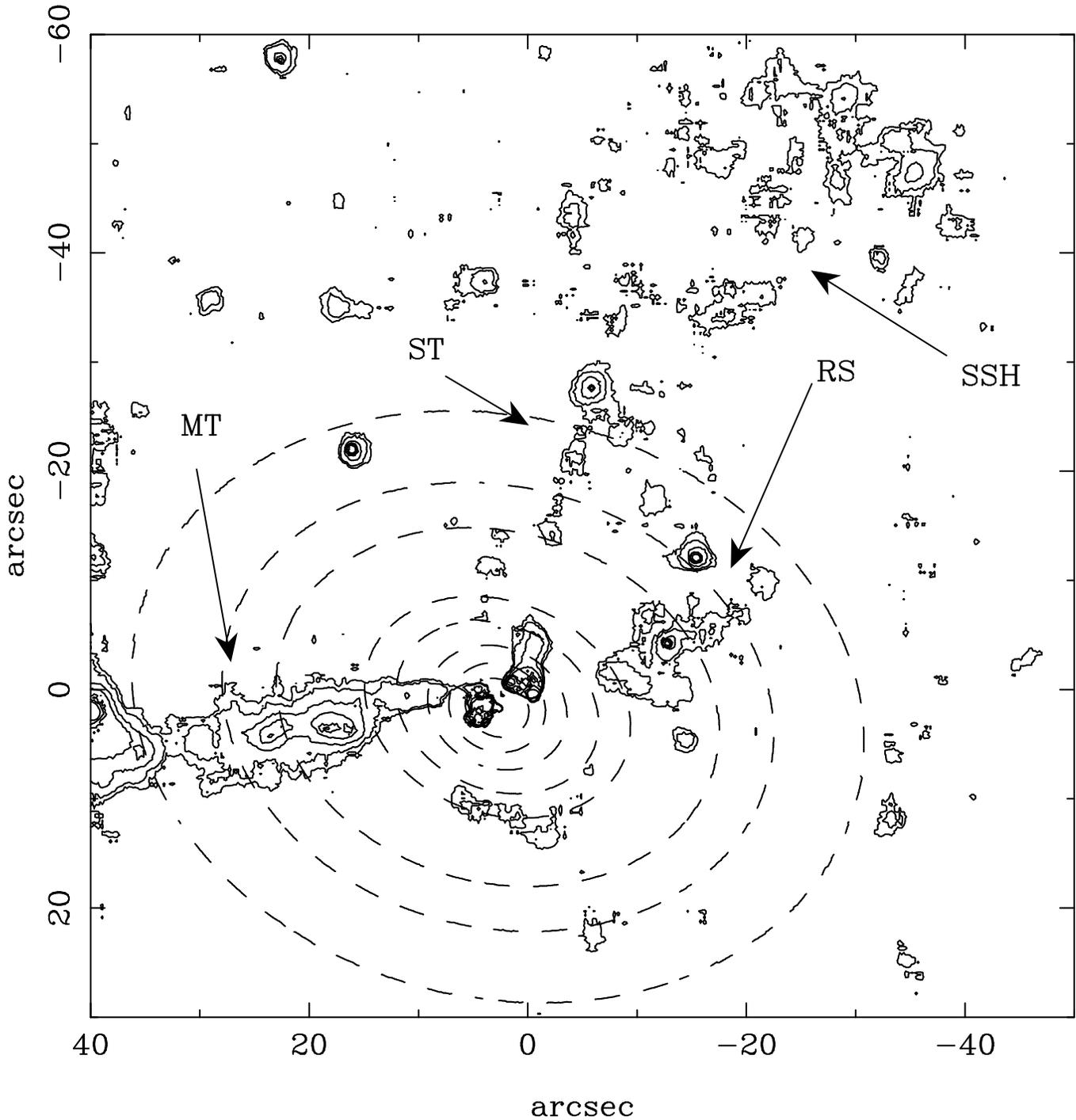}
\caption{Model-subtracted isophotal image of IC~1182 in Gunn~r band. The dotted lines are the isophotes of the model galaxy. Structures like the Main Tail (MT), but also a complex central region, a Second Tail, labeled ST, and
diffuse material (SSH and RS) are revealed}
\label{modelsub}
\end{figure*}

To enhance further the fainter details, we have used the higher resolution pictures obtained in 1995 and built a model galaxy as explained before. The model subtracted image in the Gunn~r band is shown in the Figure~\ref{modelsub}. Several structures are clearly distinguished. First, the linear feature we just mentioned, seemingly starting at the center of the galaxy and running northward for 38\arcsec~ (27 kpc), at PA  = 342, is well visible. We call it the Second Tail (ST), and it is reported here for the first time. It also presents a knotty structure, even if the condensations are not too prominent. Since the image corresponds to the Gunn~r band, it would reflect the H$\alpha$ luminosity of the knots, if they are emitters. We anticipate here that they were not detected in the narrow band H$\alpha$ images. This tail ends with a diffuse, S-shaped structure (SSH), visible toward the NW side. It is very faint and contains three condensations (clearly visible also in the insert in Figure~\ref{Bfilter}). It has a dimension of 41\arcsec, corresponding to 29 kpc, and reaches more than 50 kpc from the central region. This is roughly the same length as that of the previously known tail. Finally there is also a residual structure (RS), marked in the  figure, opposite to the central knots with respect to the dust feature we have reported.

\begin{figure*}
\centering
\includegraphics[angle=-90,width=\textwidth]{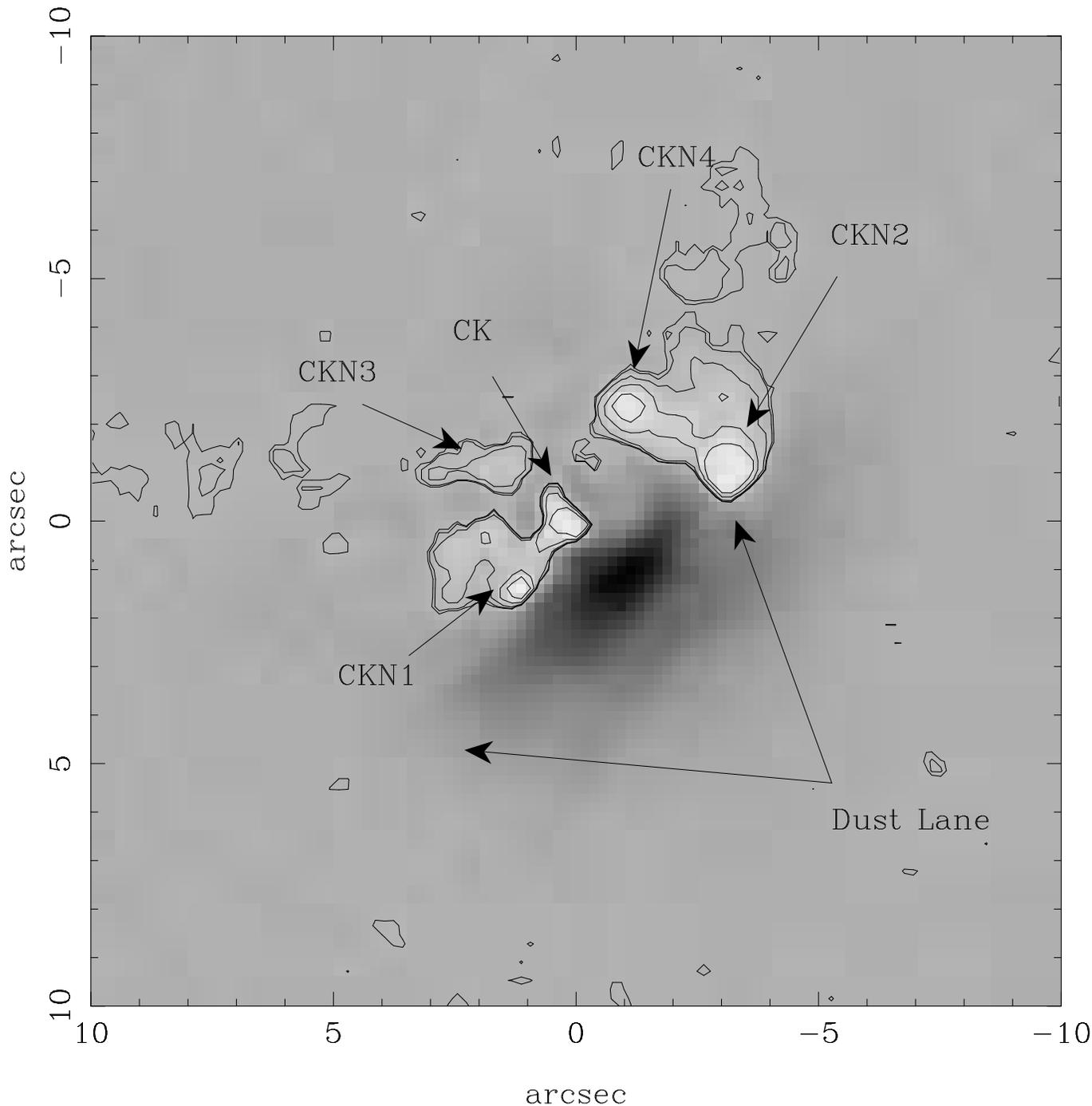}
\caption{Isophotal tracing of the central region of the model subtracted Gunn~r band image of IC~1182}
\label{centralknots}
\end{figure*}

The details of the central region of IC~1182 revealed in the model subtracted image are presented in the Figure~\ref{centralknots}. The absorption structure we mentioned earlier is clearly visible. It is elongated in the NW-SE direction and has a size of about 7\farcs5, corresponding to $\approx$ 5 kpc. There are also 5 condensations, labeled CK and CKN1 to CKN4, clearly detected. Three of them are along the  border of the dust structure. The brightest central knot  is CK. As we will discuss below, it is not exactly the center of the continuum light distribution, that appears to be displaced to the North by 142 pc. It is the head of an elongated structure at about PA = 135, that includes CKN1. The knots CK and CKN2, at 3\farcs43 correspond to those reported by Rafanelli et al. (1999), whereas the remainder are reported here for the first time. It is interesting to notice that CKN3, which is located very close to the galaxy center, is elongated in the direction of the eastward tail and, in fact, could be part of it. All the knots are well resolved in the broad band images, with typical sizes over 1 kpc. 

Otherwise, the most prominent optical characteristic of IC~1182 is indeed the presence of a linear, knotty structure, that seems to start in the central region of the galaxy and extends 88\arcsec ~to the East. It has been also detected in HI (van Driel at al. 2003), who reported observations of the structure and suggested that the bright knot at its tip could be a dwarf galaxy. Looking at the aspect ratio and faint continuum luminosity of the linear structure, it appears to satisfy the most important conditions used by Keel (1985) to differentiate between tidal tails and optical jets. It also seems to start at the center of the galaxy and, in fact, it is rather similar to the jets seen in M87 and 3C273 for example. However, it contains stellar material, HI, and belongs to a system that appears to be the result of the merging of two galaxies, where tails are expected. The tidal nature of this long structure is strongly supported by our discovery of a second tail, as expected in such cases. Except for the difference in continuum luminosity between both tails, IC~1182 is morphologically similar to NGC~7252, the prototype of mergers (Schweitzer 1982). Consequently, we find the designation ``tails'' more appropriate than ``jets''.

This long tail, called here the Main Tail (MT; see Figure~\ref{Bfilter}~and~\ref{Iongas}), 63 kpc long, has an irregular structure, completely dominated by several resolved knots, labeled MTK1 to MTK6 in the figure. The luminosity of the knots is largely due to the presence of emission lines (see below). MTK1 and MTK2 are elongated in the direction of the tail, at PA = 97, at 14\farcs5 and 19\farcs6 respectively from the photometric center of the galaxy. Both condensations are in fact double. The intermediate region of the tail is diffuse, without prominent knots except for MTK3, itself fainter than the other. (We recall here that the bright and compact object in that region, labeled {\sl c} by Stockton 1972, is in fact a star, as confirmed by Bothun et al. 1981). The three farthest knots, MTK4, MTK5 (knot {\sl d} in Stockton) and MTK6 (knot {\sl e} in Stockton), are very bright, and all well resolved (MTK5 and MTK6 include both two bright sub-condensations at least). These three knots lie still at PA$\sim$97 with respect to CK.

\section{Structure and distribution of the ionized gas}

\begin{figure*}
\centering
\includegraphics[angle=-90,width=\textwidth]{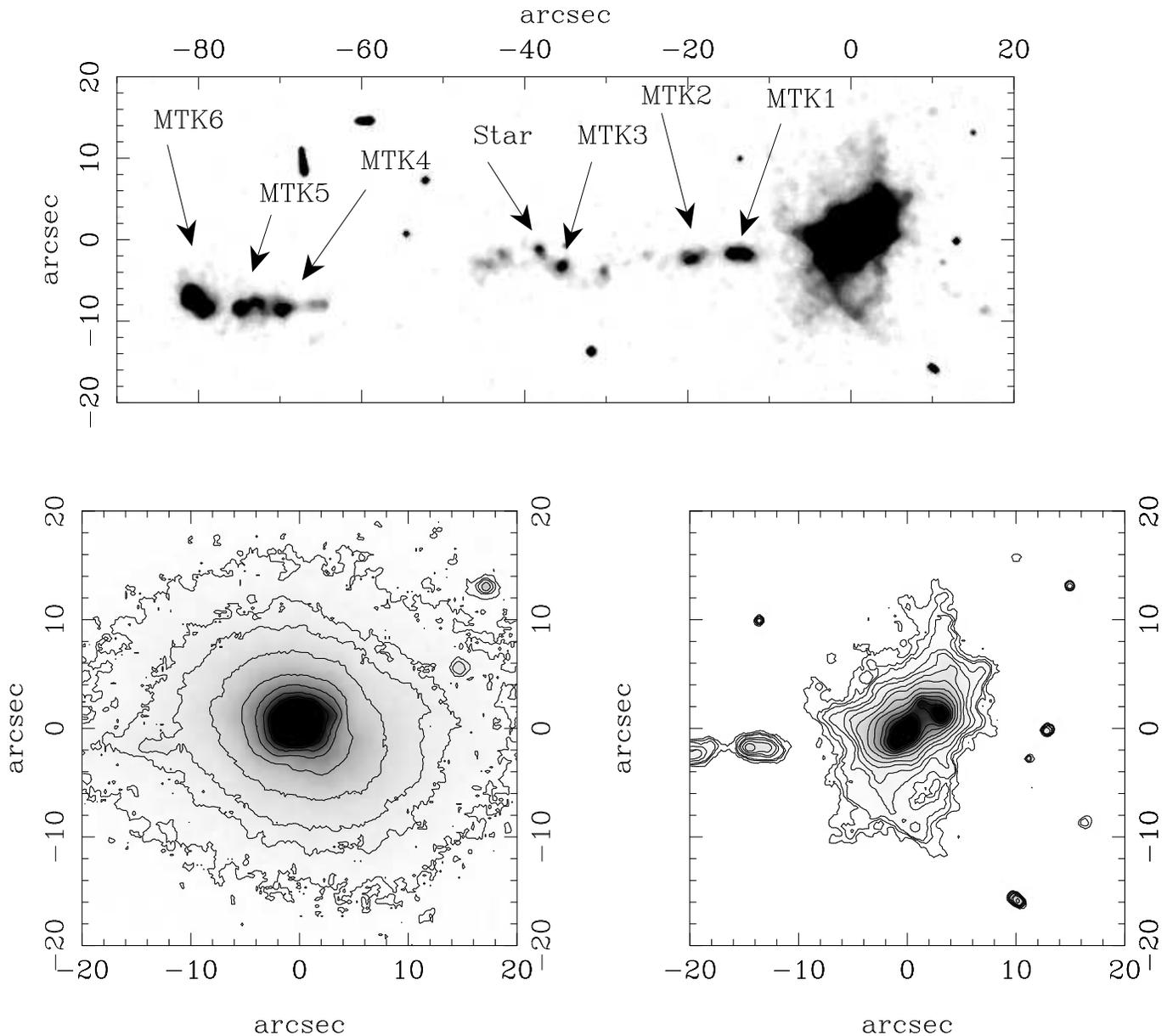}
\caption{The distribution of the ionized gas in IC~1182 as derived from the H$\alpha$ light distribution. The upper image reveals its extension and large scale structure, in particular the structure of the Main Tail. The two bottom images correspond to the central region of the galaxy. Three distinct condensations are seen in H$\alpha$ (bottom right) in the central body: CK (with CKN1), CKN2, and CKN4, and the innermost condensations of the MT. The continuum image (bottom left), shows a much more smooth structure without condensations. Notice the difference between the PA of the major axes of the continuum and emission lines light distributions}
\label{Iongas}
\end{figure*}

The H$\alpha$ line and continuum flux distributions are shown in the Figure~\ref{Iongas}. The image in the upper panel reveals the size of the emitting gas, with faint, diffuse extensions emerging in the form of plumes and arcs from the main body of the galaxy. In particular, a remarkable arc-like structure emerges toward the South and reenters the galaxy from the West, with a size of 2~kpc. 
%

The two bottom panels correspond to the central region of the galaxy. Several condensations are distinctly seen in H$\alpha$ (right bottom panel), in coincidence with the analysis of the Gunn~r band image discussed before. The most prominent knots are CK, the central one, and CKN1 to CKN4. The brightest is CK. It shows an extension toward the SE, almost resolved, with CKN1 at the end, as already seen in the B-Band image. We have measured the positions of the knots in the line and continuum images to ascertain whether they are the same in all the bands. We find that the positions agree within 0.1 pixels, except in the case of CK, the H$\alpha$ peak is offset by 0\farcs2 (142 pc) to the South of the continuum peak of CK. The central knots are very luminous in H$\alpha$, with core luminosities, L$_{H\alpha}$, well over 10$^{40}$ erg s$^{-1}$ even before any correction for internal absorption. Their nature will be discussed in section 6.

The [OIII] image (not shown here) shows a very similar structure to that of the H$\alpha$ image for the central region. The knots are all very bright, with CKN2 the highest excitation nodule (see below). The [OIII] image shows the presence of emitting gas far beyond the region of the central knots, but it does not show the filamentary structure seen in $H\alpha$. On the other hand, the continuum image reveals a hint of the presence of the knots CKN2 and CKN4. It is true that, as explained before, the [NI]$\lambda$5199\AA~ enters the filter, and would contribute to the measured flux through the continuum filter, but at a level of about 1\%. It rather corresponds to the presence of an old stellar population associated with the emission line regions, as appreciated in the corresponding spectra. The [OIII] continuum  image looks very much the same as that in the $H\alpha$ continuum. Finally, the [SII] image also shows a distribution similar to that of H$\alpha$.

The distribution of the ionized gas in the body of the galaxy has an overall morphology that corresponds to an inclined disk, where the knots are embedded (see Figure~\ref{Iongas}). We find that the external isophotes can be fitted  with an ellipse of 17\arcsec major axis, and 6\farcs2 minor axis (12.1$\times$4.4 kpc), with an inclination of 68\degr. The center of the fitted ellipse is very close to CK. Given the uncertainties, our measurements
are compatible with CK being the center of the disk, itself at 142 pc from the center of the continuum, as already mentioned. The major axis has a position angle of PA = 120, about 10\degr~ out of the line from CK to CKN2, and about 40\degr~ from the major axis of the continuum light distribution.

The Main Tail appears very conspicuously in H$\alpha$, with several condensations, already detected in the B-band image and labeled MTK1 to MTK6. They have H$\alpha$ luminosities over 10$^{39}$ erg s${^-1}$, the brightest being MTK6, at the tip of the tail. They are brighter than most of the giant HII regions in nearby spiral galaxies (Melnick et al. 1987). Some of them like  MTK1 and MTK6 are clearly double, others, like MTK5 or MTK2 have distinct extensions. The knots, as judged from their FWHM values in the corresponding images, are well resolved (the image resolution is 0\farcs85) in all cases. The sizes range from $\approx$ 0.7 kpc to $\approx$ 1.2 kpc. In principle they could simply be complexes of smaller HII regions that have not been resolved. Otherwise, they could be proto-dwarf galaxies as those seen in the Antennae (Mirabel et al, 1992) or in Stephan's Quintet (Mendes de Oliveira et al.  2001). In fact, van  Driel et al. (2003) have recently reported that the brightest knot, MTK6, could be a tidal dwarf galaxy.

The redshift and excitation of the knots were measured from the MOS spectra. All the relevant parameters are collected in
the Table~\ref{knotsHa}.

The knots in the newly discovered tail, ST, were not detected in H$\alpha$. This could be due to insufficient resolution  of our H$\alpha$ images, or just that they are not H$\alpha$ emitters. We are inclined to accept the first hypothesis since, as discussed before, the tail was only detected in the best seeing images. In the emission line image we detected a bright region close to the nucleus, not far from the region where ST enters the central part of the galaxy. However, we think that this is not directly related to ST since its PA and position do not correspond to ST as it is detected in the broad band images.
%
\begin{table}
      \caption[]{Main properties of the H$\alpha$ emission of the central and MT knots in IC~1182$^a$}
      \label{knotsHa}
$$
      \begin{array}{lcccc}
      \hline
         \noalign{\smallskip}
Knot & cz      & [OIII]/H\beta & FWHM & {\rm L_{H\alpha}}  \\
     & (\kms)  &               &  (kpc) & x10^{40} erg s^{-1}   \\
       \noalign{\smallskip}
       \hline
      \noalign{\smallskip}
{\rm CK}     &  10300 & 2.75 & 1.2 & 123.00 \\
{\rm CKN2}    &  10400 & 4.00 & 0.8 & 26.40 \\
{\rm CKN4}    &        &      & 0.7 &  4.57 \\
\hline
{\rm MTK1^b} & 10241 & 2.48 & 0.8 & 0.34  \\
{\rm MTK2^c} & 10303 & 0.95 & 2.7 & 0.32  \\
{\rm MTK3^b} &       &      & 1.0 & 0.20  \\
{\rm MTK4}   & 10332 & 2.39 & 1.0 & 0.22  \\
{\rm MTK5}   & 10394 & 2.23 & 0.9 & 0.23  \\
{\rm MTK6^d} & 10425 & 2.23 & 1.0 & 0.74  \\
     \noalign{\smallskip}
           \hline

        \end{array}
$$
\begin{list}{}{}
\item[a] The H$\alpha$ luminosity corresponds to the cores of the knots, defined by the isophote corresponding to the FWHM. The luminosity is corrected only for external extinction for the MT knots, and for internal and external extinction for the central knots.
\item[b] Double knot. The size is for each subcondensation. The size at the FWHM of the whole knot is 1.4 kpc. The luminosity is for the whole region.
\item[c] MTK2 is diffuse and complex.
\item[d] The data are for the brightest, NE condensation.
\end{list}
\end{table}

\section {The complex kinematics of IC~1182}

\begin{figure*}
\centering
\includegraphics[angle=-90,width=\textwidth]{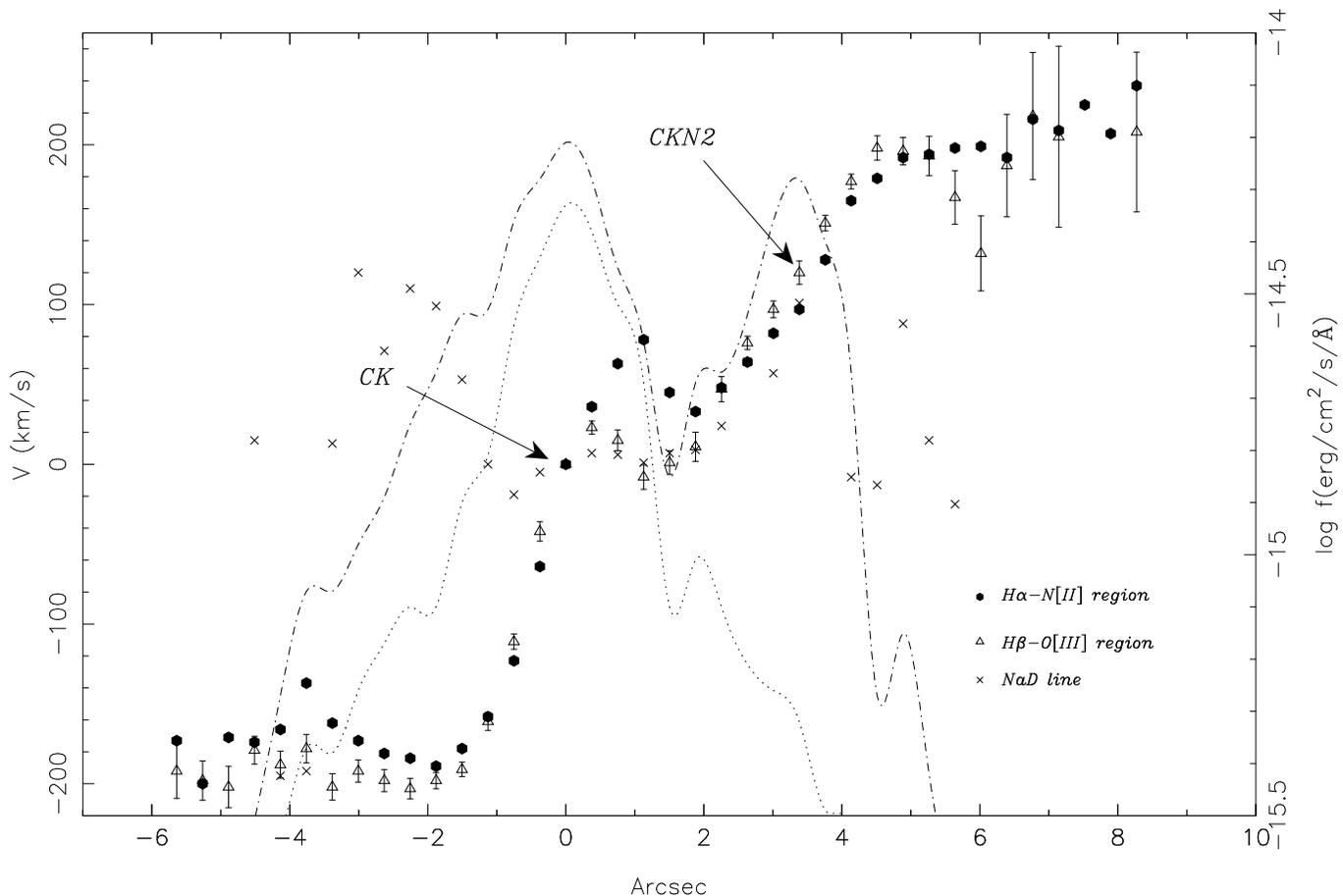}
\caption{The rotation curve measured at PA = 128, from the H$\alpha$-N[II] region ({\sl diamonds}) and from the H$\beta$-[OIII] region ({\sl triangles}). The 1-$\sigma$ error bars are plotted. We also show the velocity distribution from the NaD line ({\sl crosses}). The dashed-dotted and dotted lines correspond to the H$\alpha$ and continuum distributions respectively, along the slit (scale on the right vertical axis). The continuum flux has been multiplied by a factor of 40 to show it in the same scale}
\label{rotacion}
\end{figure*}

The spectra across CK and CKN2, at PA = 128\degr, show a peculiar rotation curve of the ionized gas. We notice that the rotation curves obtained from the H$\alpha$- [NII] region of the two NOT spectra taken with grisms \#7 and \#8, are in perfect agreement. The bluer grism \#7 spectrum also includes the H$\beta$ and [OIII] lines, that were used to determine the corresponding rotation curve. As illustrated in Figure~\ref{rotacion}, there are marked differences between the two rotation curves derived from the H$\beta$-[OIII] and H$\alpha$-[NII] regions. Given the agreement we have found between the rotation curves derived from the H$\alpha$-[NII] region in both spectra, the differences between the red and blue rotation curves have to be considered as relevant.
%
%

Indeed, these differences and the overall structure of the measured velocity distribution along PA = 128 can be understood in terms of two different rotating systems. The largest would be centered in CK, reaching the flat region of the rotation at $\sim2$\arcsec, with an   observed amplitude of 200\kms. The systemic redshift corresponds to 10300 \kms. The second system, smaller, would have the kinematic center, within our resolution, at CKN2, reaching the flat regime at $\sim 1\farcs5$ from CKN2, with an observed amplitude of $\sim$100 \kms. The relative velocity between the two centers amounts to $\sim$100 \kms. Given that the NOT spectra do not have enough resolution to allow the simultaneous detection of the two components, the position of the center of the observed spectral lines would reflect the relative weight of the components, producing the kind of data we obtain. 

\begin{figure*}
\centering
\includegraphics[angle=-90,width=\textwidth]{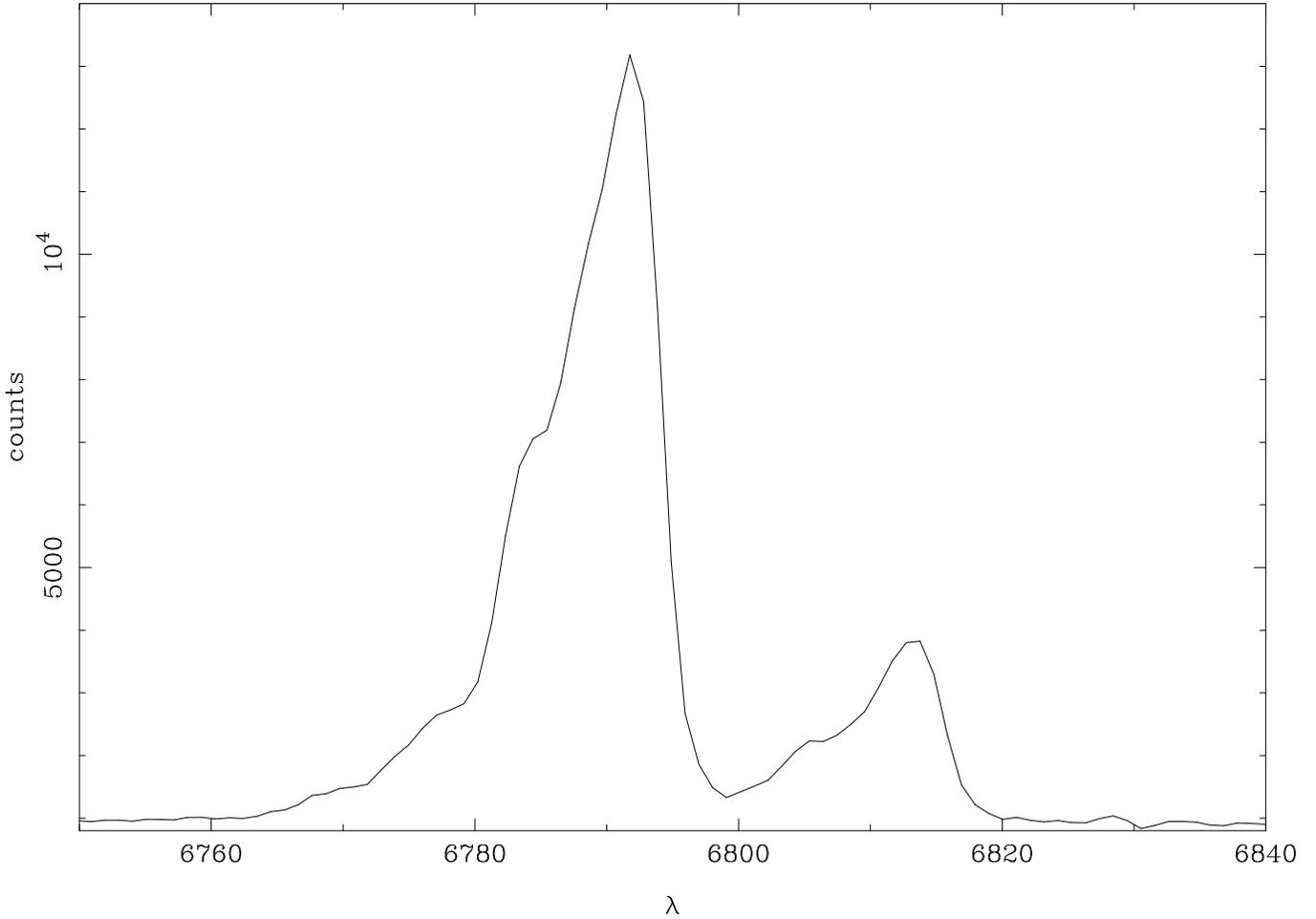}
\caption{Tracing of the H$\alpha$-[NII ]region of the spectrum of the central section of the data taken with the VLT through CKN2 and CKN4, to show the structure of the lines. }
\label{perfiles}
\end{figure*}

\begin{figure*}
\centering
\includegraphics[angle=-90,width=\textwidth]{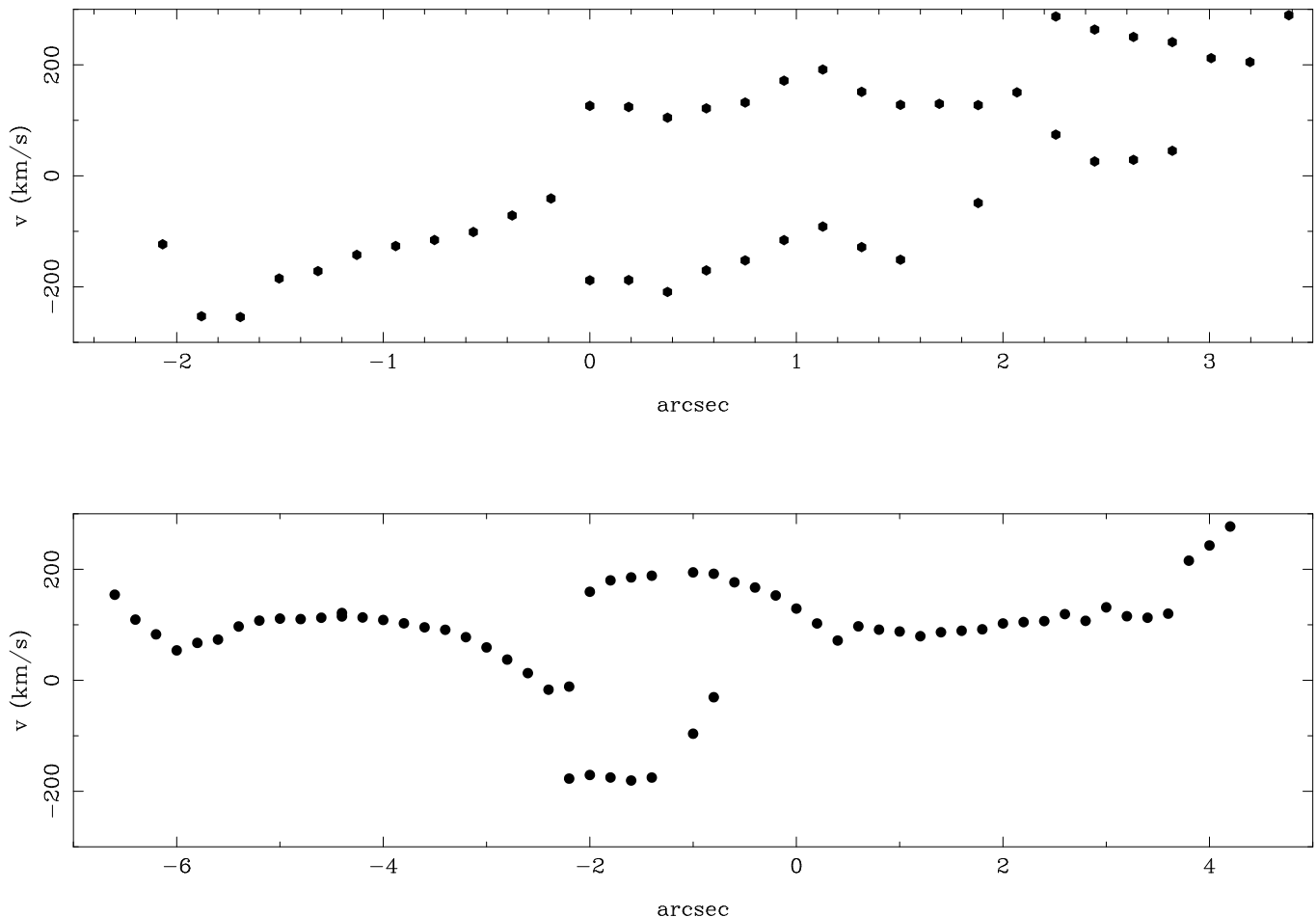}
\caption{The upper panel corresponds to the NOT grism \#13 velocity data from the [OIII] and H$\beta$ lines, measured along the slit at PA = 113, through the center of the galaxy. The lower panel presents the velocity data, measured with the H$\alpha$ and [NII] lines, from the VLT spectrum along the slit at PA = 62, through the knots CKN2 and CKN4. The positions are referred to CK and CKN2 respectively, and the velocities are given relative to the systemic measured velocity of 10300 \kms. The 1-$\sigma$ errors are of the size of the symbols,}
\label{highres}
\end{figure*}

The higher resolution data are consistent with the previous considerations. The VLT data were taken with the slit at PA = 62, i.e., 50\degr~ from the major axis we have deduced for the rotating disk, and through the knots CKN2 and CKN4, i. e., at more than 3\arcsec~ from CK. The lines are observed to be double peaked (Figure~\ref{perfiles}). 
%
\noindent
The small amplitude system is detected only in a very small region, whereas the larger amplitude system shows a structure as expected for the positioning of the slit (Figure~\ref{highres}). It is worth to notice that the NaD line (not shown), coming mainly from the region around CK (brighter by a factor of 10 than the region around CKN2 in the  continuum), does not show any rotation pattern in any of the spectra. The NOT spectrum taken with the slit at PA = 113, and grism \#13, shows [OIII]$\lambda$5007 and $H_{\beta}$. Both emission lines are also double peaked. The main system is traced only in the central 4\arcsec. The data agree with those from the lower resolution spectra. The smaller system is only detected within 1\arcsec, west side from CK (see Figure~\ref{highres}).
%

The data from all the spectra are consistent with the presence of two rotating systems. The main one would have a major axis position angle consistent with that of the distribution of the ionized gas, i. e., PA = 120. The amplitude, once  corrected for the inclination of i =  68\degr, amounts to $\sim230$km s$^{-1}$. This organized rotation corresponds to a massive galaxy with an indicative central mass (point like model) $M\sim 2\times$10$^{10}$ M$_{\odot}$. The data also indicate that the stellar component is not rotating. We have measured a velocity dispersion of $\sigma$ = 200 km s$^{-1}$. The second system, around CKN2, shows also an organized rotation, with an observed amplitude of 100 \kms, PA = 128.

\section{The line ratios: diagnostics, reddening and metallicity}

The line intensities were measured in all sections where they are well detected in the NOT spectra. The H$\alpha$ and continuum distributions are shown in Figure~\ref{rotacion}. The line ratios along the slit are plotted in Figure 8. As already seen in the narrow band images, the H$\alpha$, [OIII], and [SII] emission can be traced till very far from the center of the galaxy. Even fainter lines as [OI] are detected at distances of several kpc from the center. 
%
\begin{figure*}
\centering
\includegraphics[angle=-90,width=\textwidth]{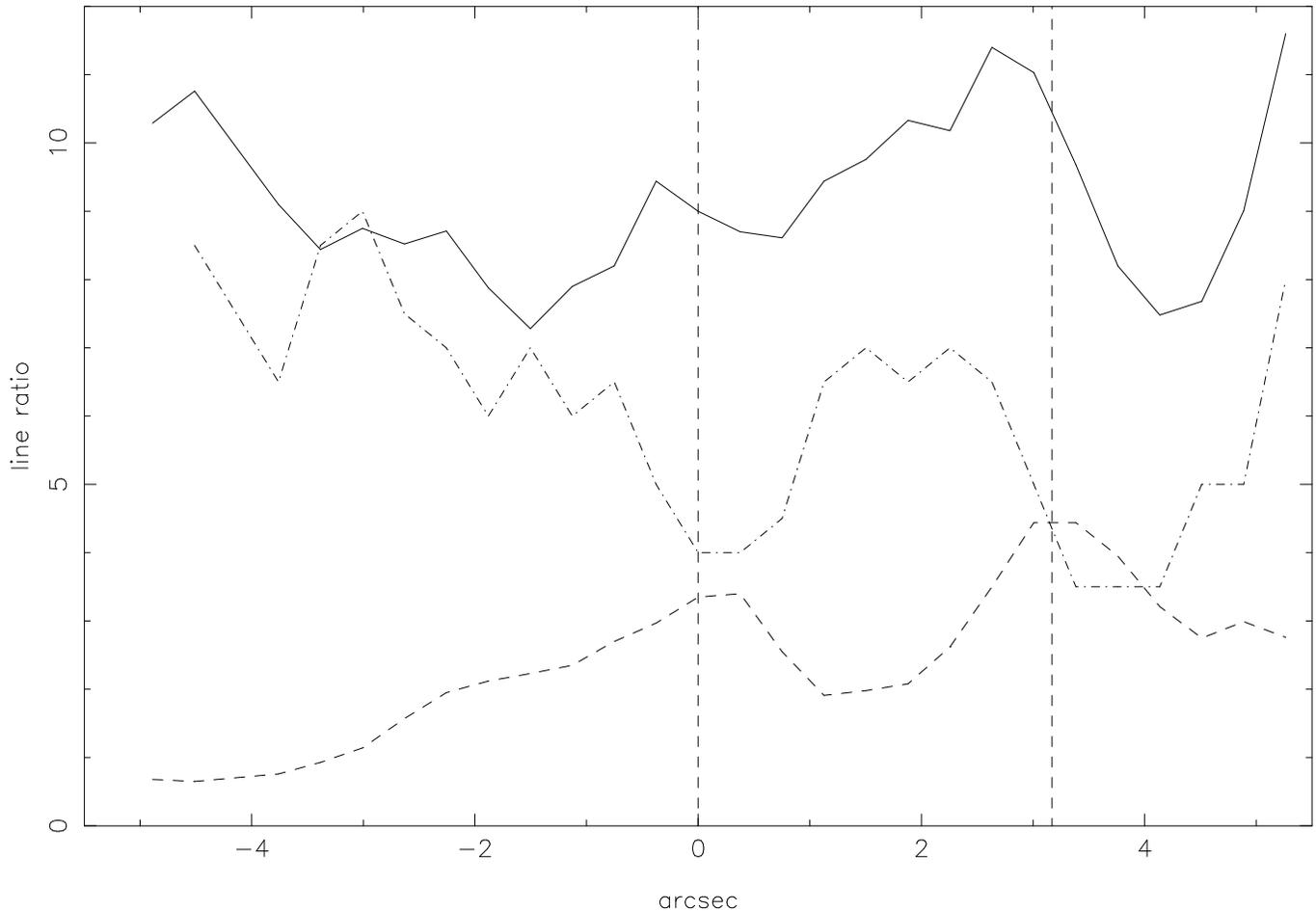}
\caption{The line ratios distribution along the slit, measured from the ALFOSC spectra at PA = 128. (a) H$\alpha$/H$\beta$ (solid line); (b) The excitation ratio [OIII]/H$\beta$ (dashed line); (c) The [OI]/H$\alpha$ ratio (dotted-dashed line). This latter ratio was multiplied by 50 to put it on the same scale as the other ratios. The vertical lines mark the position of CK and CKN2}
\label{lratios}
\end{figure*}
%

The first aspect we notice is the high H$\alpha$/H$\beta$ ratio at all positions along the slit. It was measured  always greater than 7, reaching values between 8 and 10 in the central region. In principle, that large value of the Balmer decrement could be due, at least in part, to the presence of absorption components under the Balmer lines. Indeed, that cannot be excluded before better resolution data are used and fitted by models. In any case, in the brightest regions, where we find the continuum to be very faint, and any Balmer absorption could only play a minor role, the decrement is very high. The observed range in the H$\alpha$/H$\beta$ ratio does correspond to 0.8 $\leq$ E(B$-$V $\leq$ 1.22. Taking into account the variations of the Balmer decrement along the slit, we consider a color excess of E(B$-$V) $\approx$ 1 as representative for the ionized gas component for the main body of the galaxy. Since the galactic extinction amounts to E$_{B-V}$ = 0.08 only, we conclude that IC~1182 has an important internal extinction and, therefore, other than the dust structure we have described before, IC~1182 would have important amounts of distributed dust. The typical column density toward it would be 5.8$\times$10$^{21}$ atoms/cm$^2$.

The excitation ratio, [OIII]/H$\beta$, shows two peaks, corresponding to the centers of the brightest knots, CK and CKN2. The highest value is reached at the core of CKN2, with values above 4. The excitation is important, about 2 or higher, almost all along the slit, and the [OIII] line can be traced as far as the main Balmer lines. Another remarkable aspect is the ubiquity of the [OI]$\lambda$6300 line. It appears as extended as the other lines. Its observed intensity relative to H$\alpha$ is very often $\geq$ 0.1. Those values are the most extreme that can be explained by stellar photoionization models (Evans and Dopita 1985), unless they are artificially high because of the presence of absorption under H$\alpha$. All in all, the high values of the low ionization [OI]/H$\alpha$ and  [SII]/H$\alpha$ line ratios found all along the slit, together with the rather moderate ratios of the high ionization lines, are nominally compatible with photoionization by stars with, perhaps, some underlying absorption under H$\alpha$. In any case, they would be indicative of a high effective temperature, over 40,000 K, all over the galaxy. The existence of shocks cannot be excluded, but they do not appear as a major contributor to the observed line ratios, as judged in particular by the [NII]/H$\alpha$ ratio, always smaller than 0.30.

\subsection{The Central Knot, CK}

\begin{figure*}
\centering
\includegraphics[angle=-90,width=\textwidth]{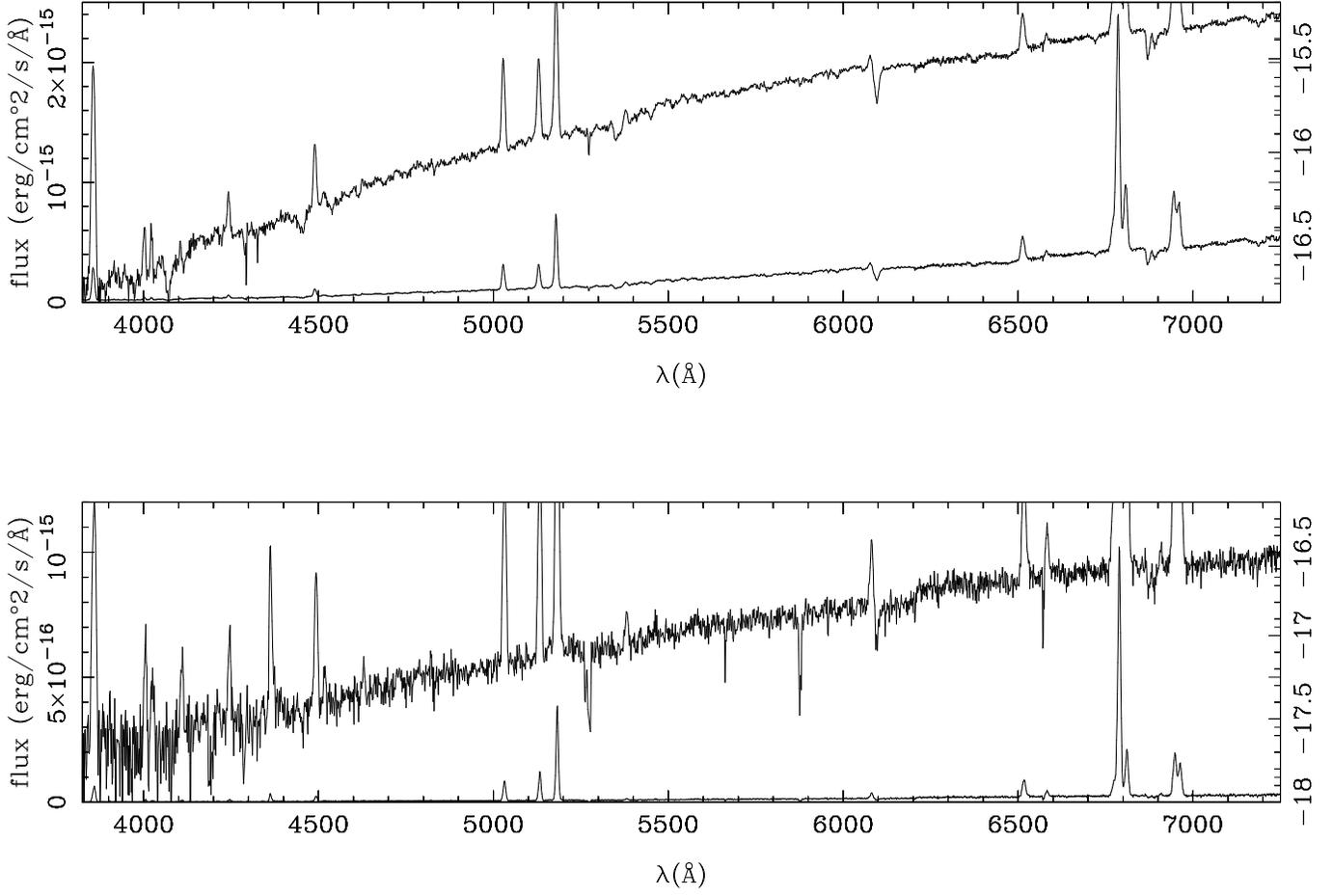}
\caption{The spectra of the most prominent knots, CK (upper panel) and CKN2 (bottom panel). For both knots the spectra are shown in linear (left) and logarithmic (right) scales}
\label{spectra}
\end{figure*}

We have shown before that CK is the center of the distribution of the ionized gas, even if it is 142~pc South from the center of the continuum, that is very faint. The emission is very compact and shows, even before any extinction correction, relatively strong high excitation lines like [NeIII]$\lambda$3868 and the [OIII]$\lambda\lambda$4363,4959,5007 lines (see Figure~\ref{spectra}). These together with the observed strength of the low ionization lines, particularly [OI] and [SII], push IC~1182 toward the region occupied by active objects in the diagnostic diagrams (Veilleux \& Osterbrock 1987).
%
%

The observed H$\alpha$/H$\beta$ ratio indicates an extinction amounting to  E(B$-$V) = 1.06 (C(H$\beta$)= 1.40). The line ratios, corrected according to this value of the extinction, are given in Table~\ref{lratios}. Plotting those line  ratios on the diagnostic diagrams by Veilleux \& Osterbrock it can be concluded that:
%
\begin{itemize}
\item{In the [NII]/H$\alpha$ {\sl versus} [OIII]/H$\beta$ diagram, CK is placed within the locus of the HII regions}
\item{In the [SII]/H$\alpha$ {\sl versus} [OIII]/H$\beta$ diagram, CK is (marginally) in the region of the active objects}
\item{In the [OI]/H$\alpha$ {\sl versus} [OIII]/H$\beta$ diagram, CK is (marginally) in the zone of the Seyfert 2 objects}
\end{itemize}
\noindent
i.e., the nucleus of IC~1182 cannot be classified as either active or HII-like without some ambiguity. We notice that the diagnostic ratios least affected by extinction indicate that CK is an HII-region like object. Besides, there are other properties to consider. First of all, as discussed before, it cannot be excluded that the Balmer line intensities  could be affected by underlying absorption. As it can be seen in Figure~\ref{spectra}, the stellar contribution to the continuum is not negligible in CK. The sense of the correction is to push IC~1182 toward the area occupied by HII regions in the diagnostic diagrams. Next to consider is the line profiles, that reflect the presence of several   components, producing a mixed spectrum.
%
\begin{table}
      \caption{Extinction corrected line ratios (in log)$^a$}
      \label{lratios}
$$
      \begin{array}{lrr}
      \hline
         \noalign{\smallskip}
    & Nucleus &  NK2   \\
           \noalign{\smallskip}
       \hline
      \noalign{\smallskip}
{\rm C_{\beta}}     &  1.49   & 1.67     \\
{\rm [OII]\lambda 3727}   &  0.54    & 0.47    \\
{\rm [NeIII]\lambda 3869} & $-$0.72 & $-$0.60  \\
{\rm [OIII]\lambda 4363}  & $-$1.21 & $-$1.40  \\
{\rm [OIII]\lambda 5007}  &  0.44   &  0.60    \\
{\rm [NI]]\lambda 5200} & $-$1.04 & $-$1.28  \\
{\rm HeI\lambda 5876}     & $-$0.11 &  $-$0.12 \\
{\rm [OI]\lambda 6300}    & $-$0.94 &  $-$1.01 \\
{\rm [NII]\lambda 6584}   & $-$0.55 &  $-$0.70 \\
{\rm [SII]\lambda\lambda 6717,6731} &  $-$0.38 &  $-$0.52  \\
{\rm [SII]\lambda 6717/ [SII]\lambda 6731} & 0.17 &  0.15  \\
\hline
{\rm n_e (cm^{-3})}    &  \approx 100 &  \approx 100 \\
{\rm T_e (K)}               & 16100     & 11600  \\
{\rm 12 + log [O/H]}        & 7.70      & 8.00   \\
 \noalign{\smallskip}
           \hline

        \end{array}
$$
\begin{list}{}{}
\item[a] The line ratios are relative to H$\beta$ till [NI]]$\lambda$5200 (included), and relative to H$\alpha$ for the other lines, except the last one which is the ratio of the two [SII] lines. All the values are from the NOT spectra
\end{list}
\end{table}
%

Similar line ratios are also found in other sites in the galaxy, which makes the idea of some kind of nuclear activity  even less attractive. The presence of low and high ionization lines throughout the galaxy is indicative of extremely  powerful and generalized star formation processes. The low densities and the normal ratios of the low ionization lines with respect to the Balmer lines is also indicative of pure photoionization by stars.
 
Accepting the HII nature of the emission region we see in IC~1182, the line ratios correspond to a very low electronic density (the sulfur lines indicate that we are in the low density regime). The electronic temperature of CK then amounts to more than 16000 K. As for the metallicity, since the O[III]$\lambda$4373 is well detected, it can be determined in the standard way. We find Z =  1/16 Z$_{\odot}$ (see Table~\ref{lratios}). The main uncertainties come form the error in the [OIII] auroral line and from the high extinction correction to apply to the O[II]$\lambda$3727 line. The fact that the O[III]$\lambda$4363~ line is detected indicates that the metallicity is well below solar. The He abundance is difficult to determine given the high extinction and the contamination by the NaD doublet. We find it lower than solar by 10-20\%.

\subsection{The knot CKN2}

The overall aspect of the spectrum of CKN2 is that of a high excitation HII region (Figure~\ref{spectra}). The continuum is less steep than that of the central region, with a smaller stellar contribution. The Balmer decrement is very high here too. It corresponds to C(H$\beta$) = 1.62, or (E(B$-$V) =  1.2 mag. The corrected line ratios (see Table~\ref{lratios}) are close to those expected for a high T$_{eff}$ ($\approx$ 50,000 K) HII region. Still, the values  of [OI]/H$\alpha$ and [SII]/$H\alpha$ place (marginally) CKN2 in the zone occupied by active, Seyfert 2 objects in the diagrams by Veilleux \& Osterbrock. As before, we notice that the best measured indicator places CKN2 in the area of the HII regions in the diagnostic diagrams, and that some absorption under the Balmer lines and/or the composite nature of  the spectrum would help to reconcile all the line ratios with those predicted by HII region models.

The electronic density of CKN2 is also low, and the electronic temperature 11600 K, significantly lower than in CK. The
O[III]$\lambda$4363 line is also detected here. The metallicity, even if almost double that of CK, is still significantly lower than solar. Finally, the He abundance is similar to that of CK, about 10\% lower than solar.

\section{Discussion and conclusions}

The global morphology of IC~1182 is reminiscent of an early type galaxy, but with important peculiarities. In the continuum light, IC~1182 shows an entangled structure, with $\varepsilon$ = 0.25 and PA = 80 when the outermost isophotes are considered. A long tail, 63 kpc long, emerges from the central region to the E, at PA = 97, with a knotty structure. We  have detected a new, much fainter tail running almost perpendicular to the previous one, and stretching for 27 kpc from the center. It is continued by a diffuse, large S-shaped structure toward the NW, already visible in the photographic picture by Arp (1972). In the model subtracted image we have constructed, one sees a large, disk-like dust structure, already hinted in the B-band image, with a size of 5~kpc. Finally, to the SW of the dust structure, there is a residual structure that appears distinctly in the model-subtracted image. These characteristics, in particular the discovery of the Second Tail, strongly suggest that IC~1182 could be a merging system. Moreover, several knots along the main tail, detected in the broad band and H$\alpha$ images, are resolved. They could be forming tidal dwarf galaxies as seen in the Antennae (Mirabel et al. 1992) or in Stephan's Quintet (Mendes de Oliveira et al. 2001), as recently suggested by van Driel et al. (2003) for the knot at the tip of MT.

The narrow band images centered in the main emission lines show a galactic size distribution of ionized gas, with plumes and filaments stretching several kpc from the central regions. The amount of ionized gas could be as high as 10$^6$ or   even 10$^7$ M$_{\odot}$. It is organized in an inclined (i =  68; PA = 120) rotation disk with a major axis of about 12 kpc.

The velocity distribution we observed at two slit positions can be interpreted as produced by two distinct rotating systems. The main system would dominate the dynamics of most of the observed ionized gas, with a rotation amplitude of $\approx$ 230 \kms. The mass responsible for that rotation would amount to 2$\times$10$^{10}$ M$_{\odot}$. The G parameter amounts to $\approx$ 300 \kms kpc$^{-1}$. In that region, the stars don't show any clear rotation pattern, and have a velocity dispersion $\sigma$ = 200 \kms. The rotation amplitude and G value are typical of the bulge of a Sb galaxy (M\'arquez  \& Moles 1999). The second system would be less massive, with an observed rotation amplitude of 100 \kms.
 
There are no clear signs of nuclear activity. The measured line ratios of the brightest knots are still compatible with photoionization by stars. We also notice that, given the complex structure of the central region, the disagreement about    the line ratios reported in earlier work, could simply reflect the differences in slit width and positioning.

Rafanelli et al. (1999) have reported rapid variability in X-rays, indicative of the presence of a very compact source. In principle this would support the idea of the existence of some kind of nuclear activity in IC~1182. This is however not confirmed by our spectroscopic analysis, that points out that all the optical properties can be explained in terms of a massive star formation process. Zezas et al. (2002) and Metz et al. (2003) have analyzed the X-ray flux distribution in the Antennae Galaxy and concluded that the X-ray emission is associated with young stellar clusters. IC~1182 could be, in that sense, similar to the Antennae Galaxy.  

The evidence of the data lead us to consider IC~1182 as an ongoing merger. The Main Tail, and the Second Tail we report here for the first time, together with the the S-shaped material observed far from the body of the system and the presence of knots in the central region are clear signs of gravitational interaction. The data we present here indicate that that merging process is now  approaching the final steps. Morphologically the galaxy is close to a spheroidal system with peculiarities, but the two galaxies are still dynamically distinct. The observed rotation  amplitudes indicate that the involved galaxies are rather different, an early and a late type spiral. The difference in luminosity between the two tails would be an indication of the mass difference between the two galaxies. The organization of the ionized gas in a disk is in agreement with the predictions by the models developed by Barnes (2002). The low metallicity deduced for the gas in the two central knots indicates that important amounts of fresh  gas have been supplied by the late galaxy in the process of interaction and merging. The SFR deduced from the H$\alpha$ luminosity is very high, amounting to 90~M$_{\odot}$ per year, after correction for the high internal extinction, which places IC~1182 among the very bright starburst galaxies.

\begin{acknowledgements}

We acknowledge enlightening discussions with Giuseppe Galletta, Isabel M\'arquez, Josefa Masegosa and Enrique P\'erez. Leif Hansen and Henning J{\o}rgensen are thanked for permitting us to use their IRAF script {\sl HJCOMP}. Bo Milvang is   thanked for taking the DFOSC spectra of IC~1182 for us. We would also thank the referee, Dr. P. Rafanelli, for his comments that helped us to improve significantly our manuscript.

MM ad JV acknowledge financial support from the Spanish Ministerio de Ciencia y Tecnolog\'{\i}a through grants PB98-0684 and PNAYA2002-01241. PK acknowledges financial support from the Danish National Science Foundation, grant no. 9701841.

\end{acknowledgements}

\end{document}